\documentclass[mypaper,8pt,twoside]{CoAst}
\usepackage{epsf,graphicx,fancyhdr}
\renewcommand{\chaptermark}[1]{\markboth{#1}{}}

\newcommand{\kopf}{\small\itshape Comm. in Asteroseismology \\ Contribution to the Proceedings of the 38$^{th}$\,LIAC\,/\,HELAS-ESTA\,/\,BAG, 2008
}

\newcommand{\Authors}[1]{\begin{center}\normalsize\bf\sf #1 \end{center}}

\renewcommand{\author}[1]{\begin{center}\normalsize\bf\sf #1 \end{center}}
\newcommand{\Address}[1]{\begin{center}\small\sf #1 \end{center}}

\DeclareGraphicsExtensions{.eps,.jpg}

\renewenvironment{abstract}{\section*{Abstract}\normalsize\sf}{}
\newcommand{\References}[1]{\begin{flushleft}{\large References\\}\vspace*{2mm}\small #1 \end{flushleft}}

\newcommand{\chapterCoAst}[2]{\chapter[\sf\normalsize #1\\ \footnotesize \hspace*{5mm}by #2 \sf\normalsize][]{#1\\}\rhead[\fancyplain{}{\sf\footnotesize \center{#1}}]{\fancyplain{}{\sffamily\thepage}}\lhead[\fancyplain{\kopf}{\sffamily\thepage}]{\fancyplain{\kopf}{\sf\footnotesize \center{#2}}}}

\renewcommand{\chaptername}{}

\newcommand{\acknowledgments}[1]{\vspace*{5mm}\noindent  \textbf{Acknowledgments.} #1}

\def\rfr{\smallskip\par\noindent
        \hangindent=7truemm
        \hangafter=1}

\begin{document}
\sf

\chapterCoAst{Ledoux's convection criterion in evolution and asteroseismology of massive stars }
{Y.\,Lebreton, J.\,Montalb{\'a}n, M.\,Godart, P.\, Morel, A.\, Noels and M.-A.\, Dupret} 

\Authors{Y. Lebreton$^{1,2}$, J. Montalb{\'a}n$^3$, M. Godart$^3$, P. Morel$^4$, A. Noels$^{3}$, M.-A. Dupret$^{5}$} 

\Address{
$^1$ Observatoire de Paris, GEPI, CNRS UMR~8111, 92195 Meudon, France\\
$^2$ IPR, Universit\'e de Rennes 1, 35042 Rennes, France\\
$^3$ Institut d'Astrophysique et de G\'{e}ophysique, University of 
Li\`{e}ge, Belgium\\
$^4$ Observatoire de la C\^ote d'Azur, CASSIOPEE, Nice, France\\
$^5$ Observatoire de Paris, LESIA, CNRS UMR~8109, 92195 Meudon, France
}

\noindent
\begin{abstract}

Saio et al.~(2006) have shown that the presence of an intermediate convective zone (ICZ) in post-main sequence models could prevent the propagation of g-modes in the radiative interior and hence avoid the corresponding radiative damping. The development of such a convective region highly depends on the structure of the star in the $\mu$-gradient region surrounding the convective core during the main sequence phase. In  particular,the development of this ICZ depends on physical processes such as mass loss, overshooting (Chiosi \& Maeder~1986, Chiosi et al. 1992, see also Godart et al., these proceedings) and convective instability criterion (Schwarzschild's or  Ledoux's criteria). In this paper we study  the consequences of adopting the Ledoux's criterion on the evolution of the convective regions in massive stars ($15$ and $20\ M_{\odot}$), and on the pulsation spectrum of these new B-type variables (also called SPBsg).
\end{abstract}


\section*{Stellar models: evolution of convective regions during main sequence and post-main sequence phases}

Stellar evolution models have been calculated with the {\small CESAM} code (Morel \& Lebreton, 2008) for masses $M=15$ and $20\ M_{\odot}$, a solar chemical composition ($Z/X=0.0245$, $Y=0.27$) without and with overshooting of convective zones ($\alpha_{\rm ov}=0.2 H_{\rm p}$). For all these models, we calculated the pre-main sequence evolution.
In models including overshooting, we assumed that $\nabla=\nabla_{\rm ad}$ in the overshoot region.

\begin{figure}[ht]
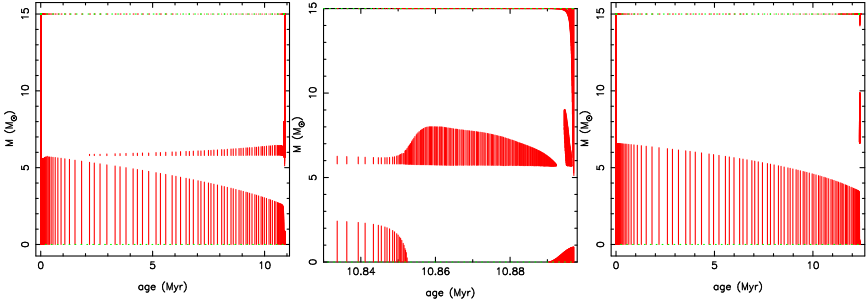

\centering
\resizebox{\hsize}{!}{\includegraphics*[angle=-90]{m15_cz_ov0.0_ms.eps}\includegraphics*[angle=-90]{m15_cz_ov0.0_pms.eps}\includegraphics*[angle=-90]{m15_cz_ov0.2.eps}
}
\caption{Evolution of the convective zones as a function of age in a $15\ M_\odot$ model calculated with the Ledoux's criterion for convection: (i) models without overshooting during the MS (left) and post-MS (center), and (ii) models with overshooting during the MS (right).}
\label{CZ}
\end{figure}

As the star evolves on the main sequence (MS), a gradient of chemical composition ($\nabla_{\mu}$) develops at the outer border of the convective core. In the context of Schwarzschild's criterion (convective instability if $\nabla_{\rm rad}\ge \nabla_{\rm ad}$), the outwards increase of opacity leads to the formation of a region of semiconvective instability outside the convective core (CC) and therefore to the mixing of matter until the neutrality of gradients is reached ($\nabla_{\rm rad}=\nabla_{\rm ad}$).  During the post-MS this region becomes an ICZ which develops as H starts burning in a shell in the $\mu$-gradient region.

\begin{figure}[ht]
\centering
\vspace*{-0.5cm}
\resizebox{\hsize}{!}{\includegraphics*[]{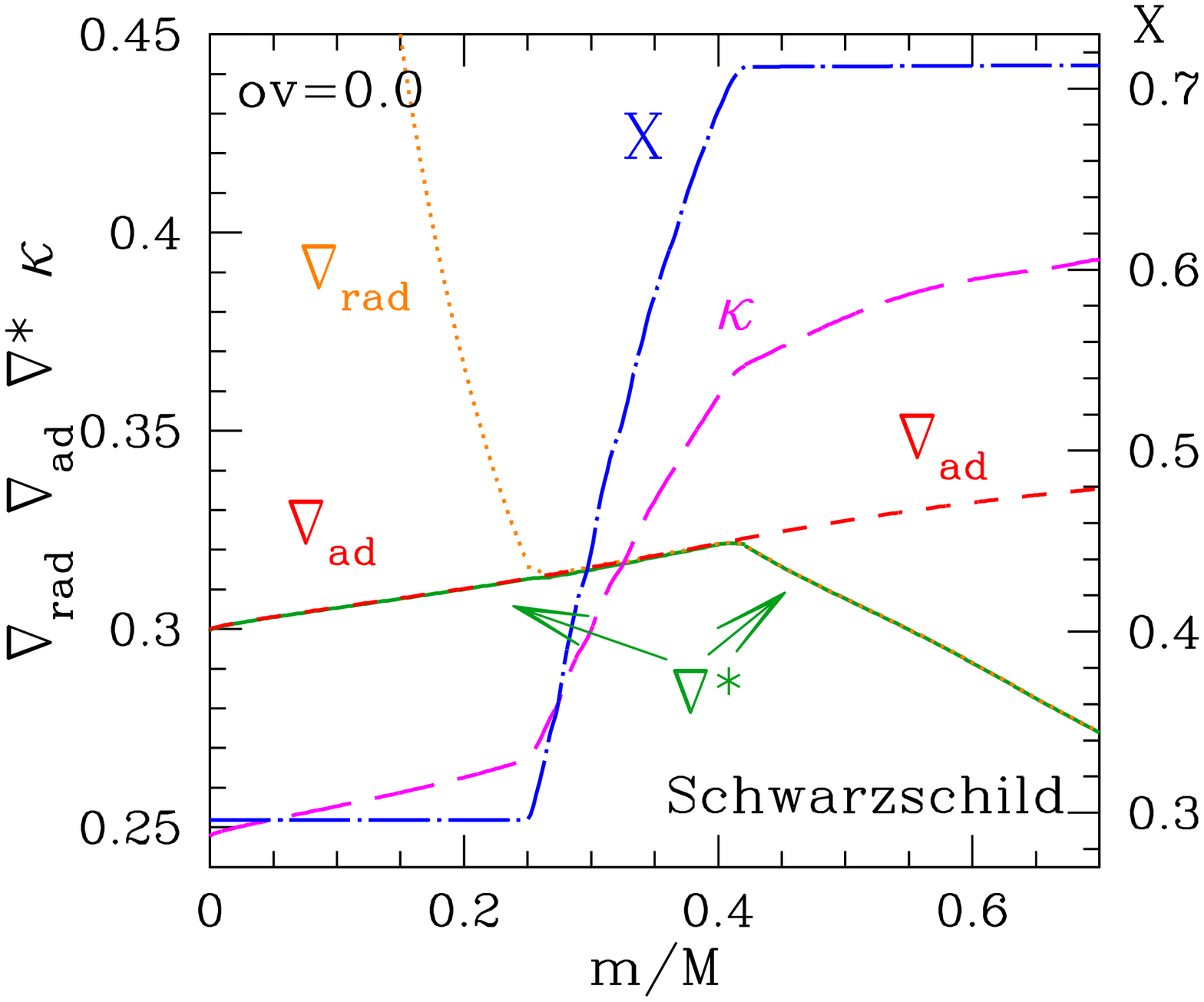}\includegraphics*[]{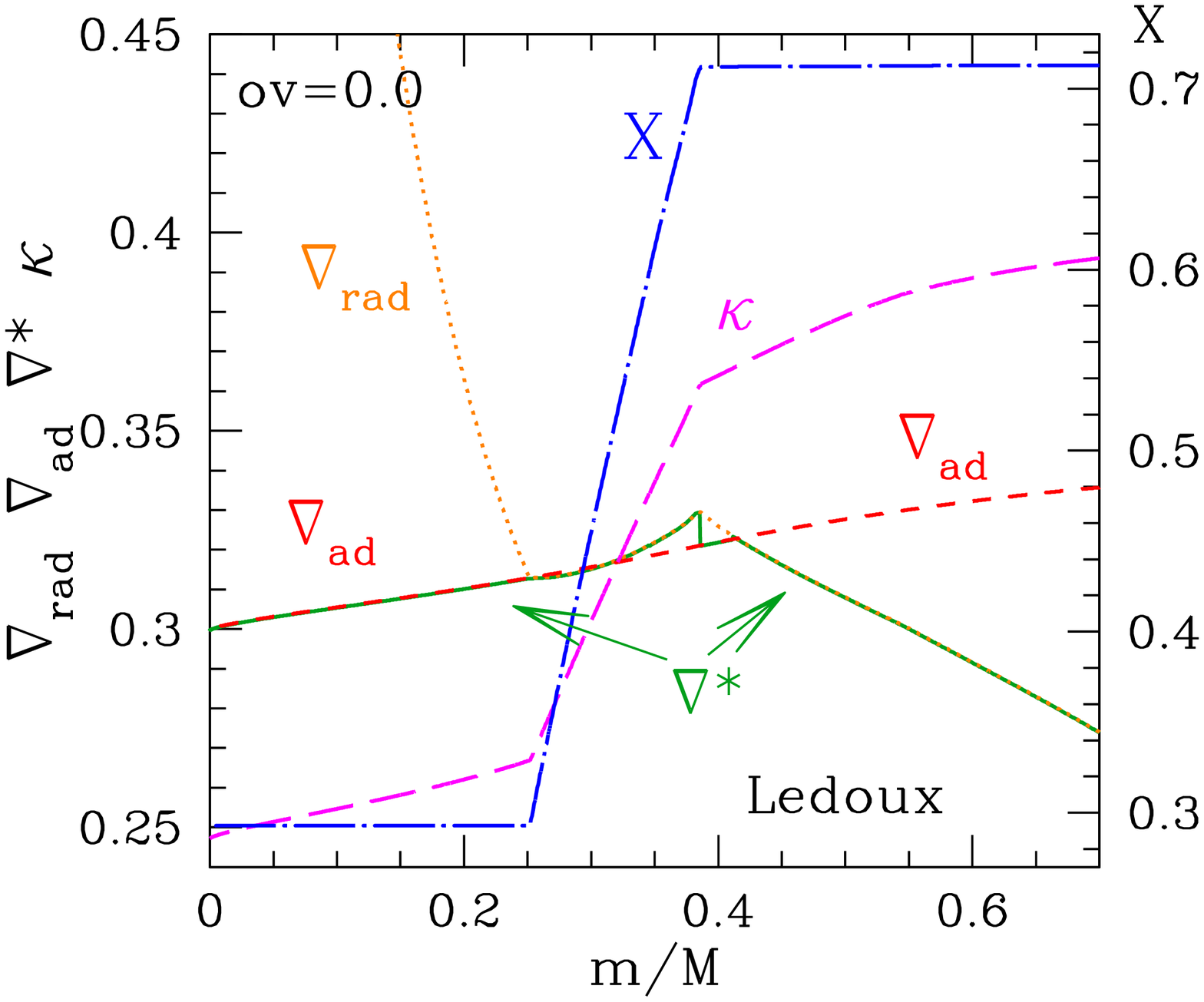}\includegraphics*[]{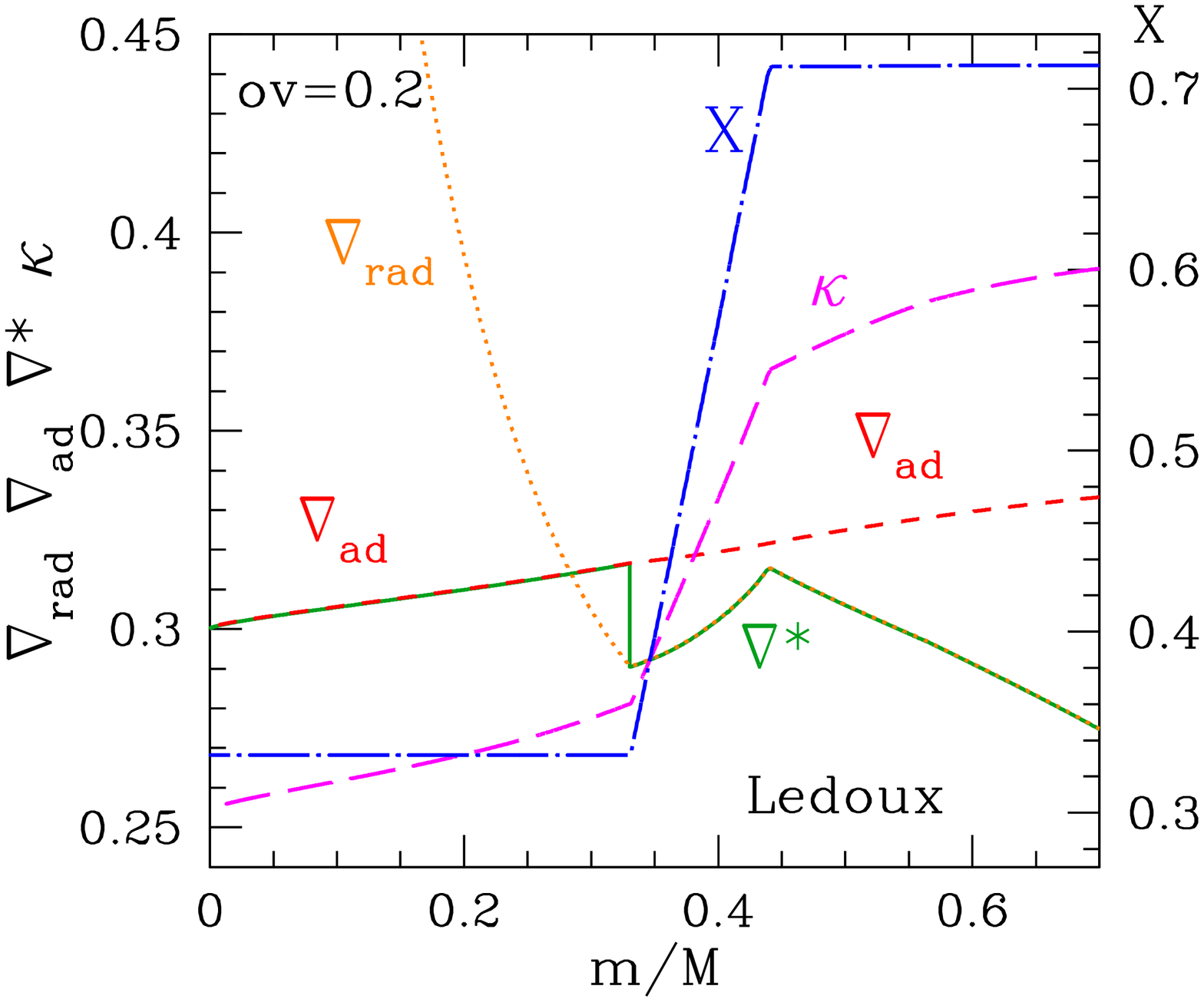}}
\caption{Temperature gradients (radiative one: dotted lines; adiabatic one: short-dashed lines; effective stellar gradient: solid lines), opacity $\kappa$ (long-dashed lines), and hydrogen profile, X (dash-dotted lines),  for a $M=15 M_\odot$ model near the middle of the MS calculated with the Schwarzschild's criterion (left) without overshooting, and with the Ledoux's criterion without (center) and with (right) overshooting.}
\label{gradients}
\end{figure}

When the Ledoux's criterion for convection (convective instability if $\nabla_{\rm rad}\ge \nabla_{\rm ad}+ \frac{\beta}{4-3\beta} \nabla_\mu$ where $\beta$ is the ratio of the gas to the total pressure) is used instead of the Schwarzschild's criterion,
the role of $\mu$-gradients  on the stability against convection is taken into account. Adopting  the Ledoux's criterion does not change the size of the CC during the MS. Nevertheless, in models based on Ledoux's criterion, a convective region located outside the CC appears during the MS phase at the base of the homogeneous region at $m/M=0.45$ in the $20\ M_\odot$ model and  $m/M=0.39$ in the $15\ M_\odot$ one  (Fig.\ref{CZ}, left panel). As can be seen in Fig.\ref{gradients} (central panel) this convective region results from the outwards increase of opacity in a region where $\nabla_{\mu}$=0 and where the Schwarzschild and Ledoux criteria are hence equivalent. The location of the ICZ  corresponds to that of the maximum extension of the CC during the MS and it remains the same during the post-MS phase until  the onset of He burning. The location and the size of the ICZ during the post-MS in models adopting the Ledoux's criterion are different from those found in models based on the Schwarzschild's criterion (see also Godart et al, these proceedings). With Ledoux's criterion, the thickness of this ICZ can reach $15-20\%$ of the total star's mass (Fig.\ref{CZ}). On the other hand, when overshooting is included, no ICZ appears during the MS for $M=15 M_\odot$  (right panels of Figs.\ref{CZ} and  \ref{gradients}) while for $M=20 M_\odot$ the ICZ appears later than in models without overshooting.

\section*{Asteroseismology of post-MS B-type stars}

We have investigated the seismic characteristics of our models according to the recipes described by Dupret et al. (these proceedings). Excitation and damping of p and g modes highly depend on the location and thickness of the ICZ, hence on the change of luminosity as the star becomes cooler. In models computed with the Ledoux's criterion, the $\mu$-gradient region located below the ICZ brings a large contribution to the Brunt-Va\"is\"ala frequency $N_{\rm BV}$ which leads to strong damping of the modes. For instance,  we find that at $\log T_{\rm eff}=4.27$ the kinetic energy of a $\ell=1$ mode in the radiative core of post-MS models of $15\ M_{\odot}$ is much higher than in $20\ M_{\odot}$ models. As a consequence, the mode in the $15\ M_\odot$ star is more damped in the radiative centre and its amplitude in the driving region at $\log T\sim 5.2$ is too low for the mode to be effectively excited.

\begin{figure}[ht]
\centering
\resizebox{\hsize}{!}{\includegraphics*[]{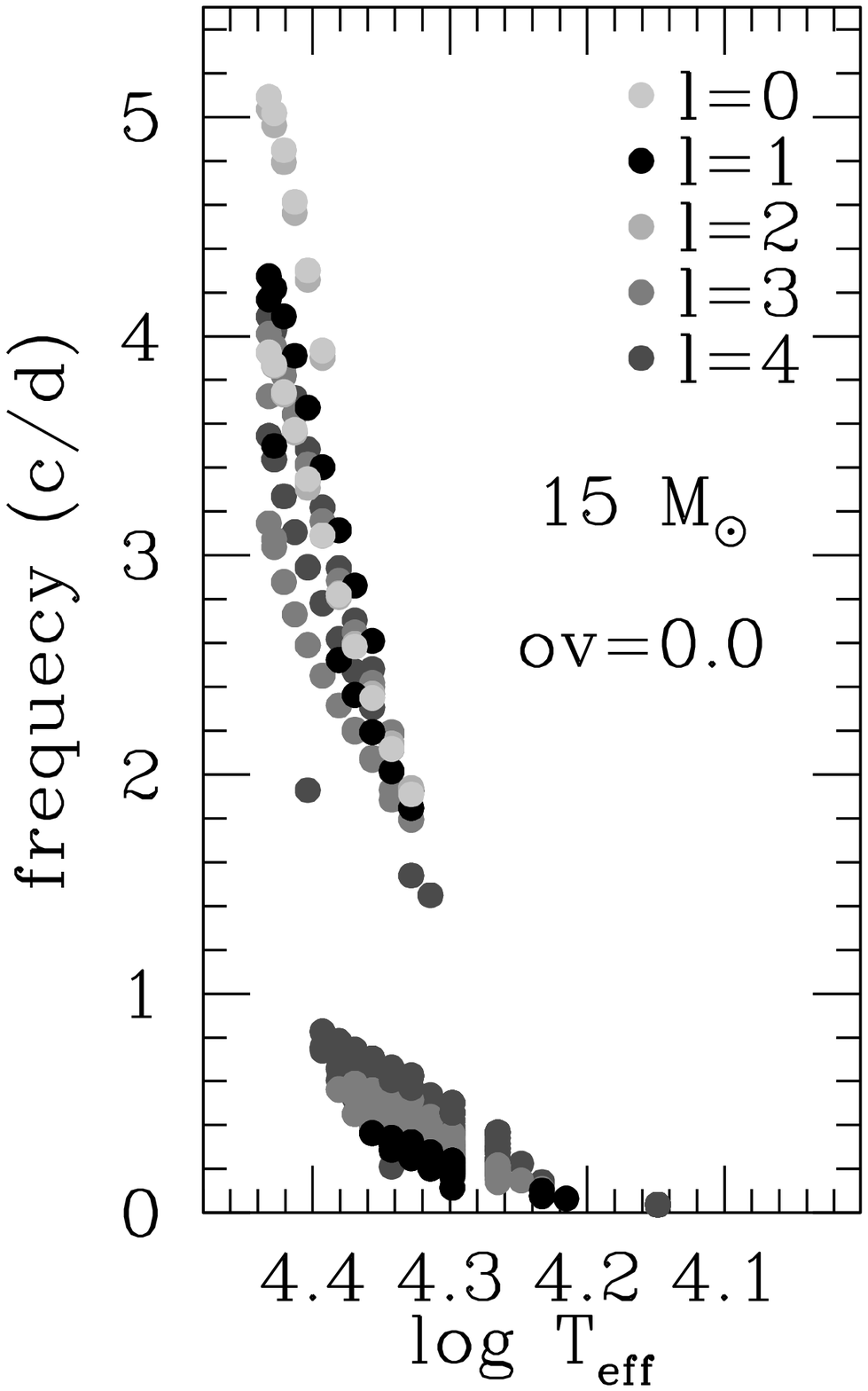}\includegraphics*[]{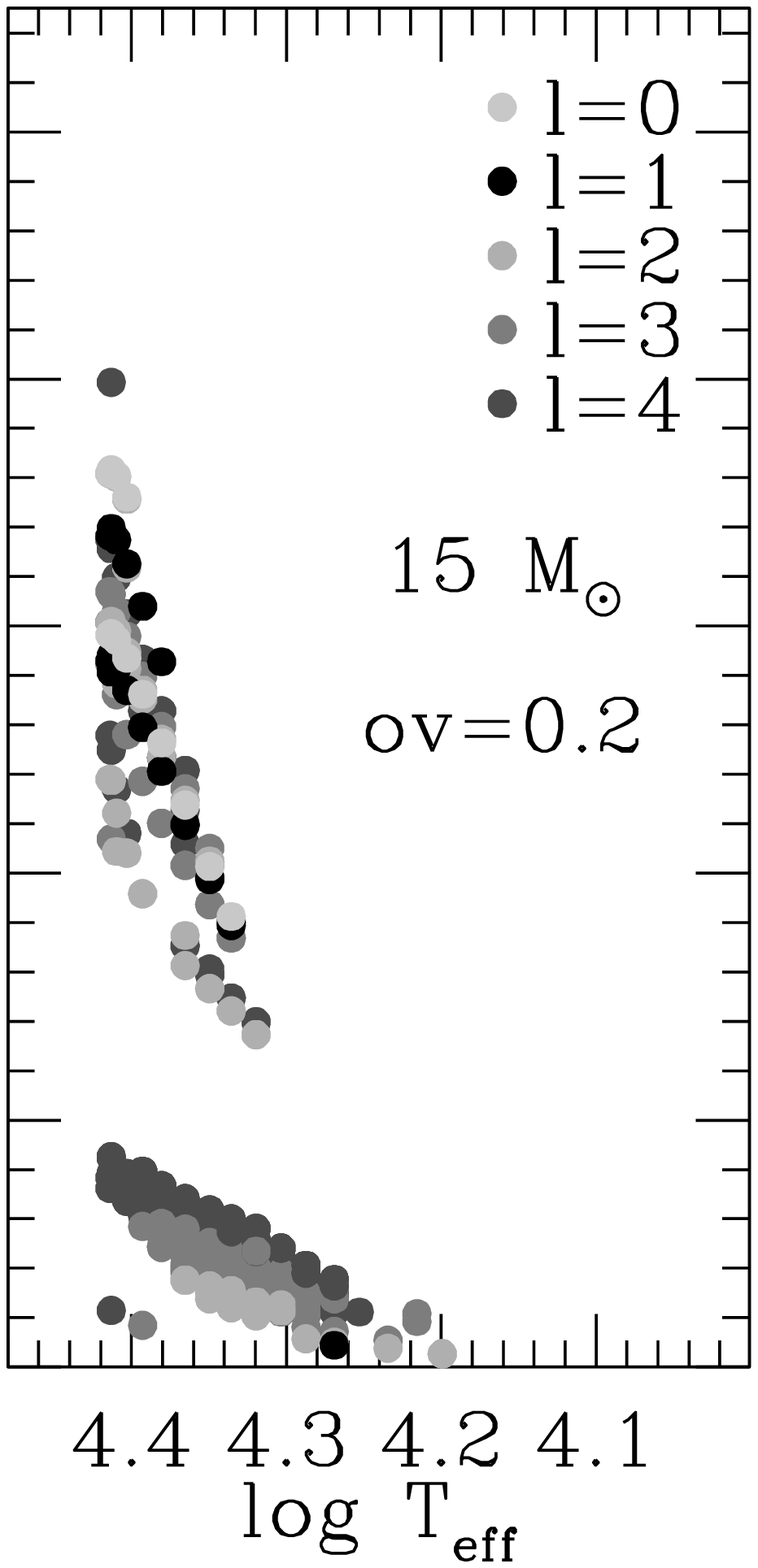}
                      \includegraphics*[]{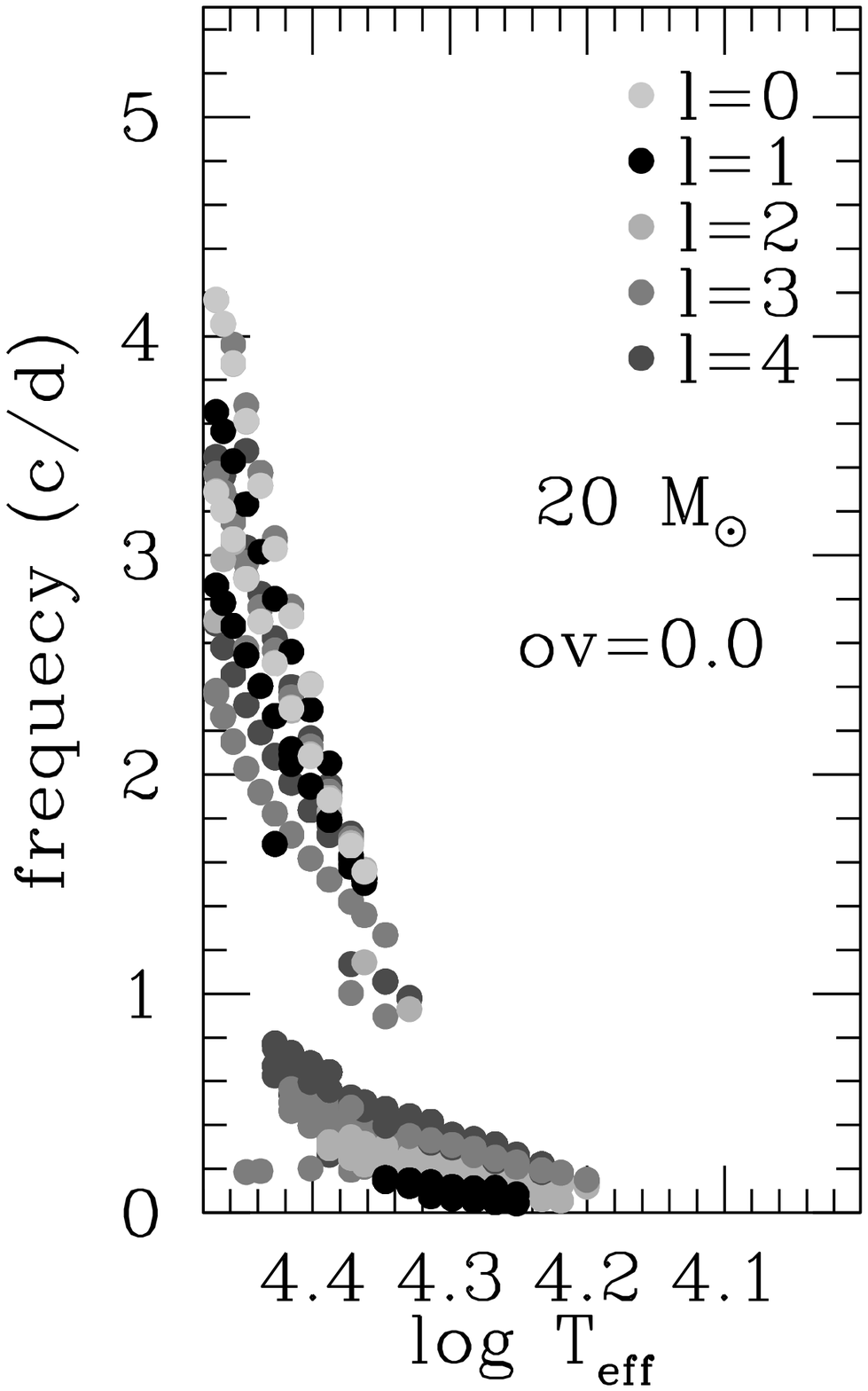}\includegraphics*[]{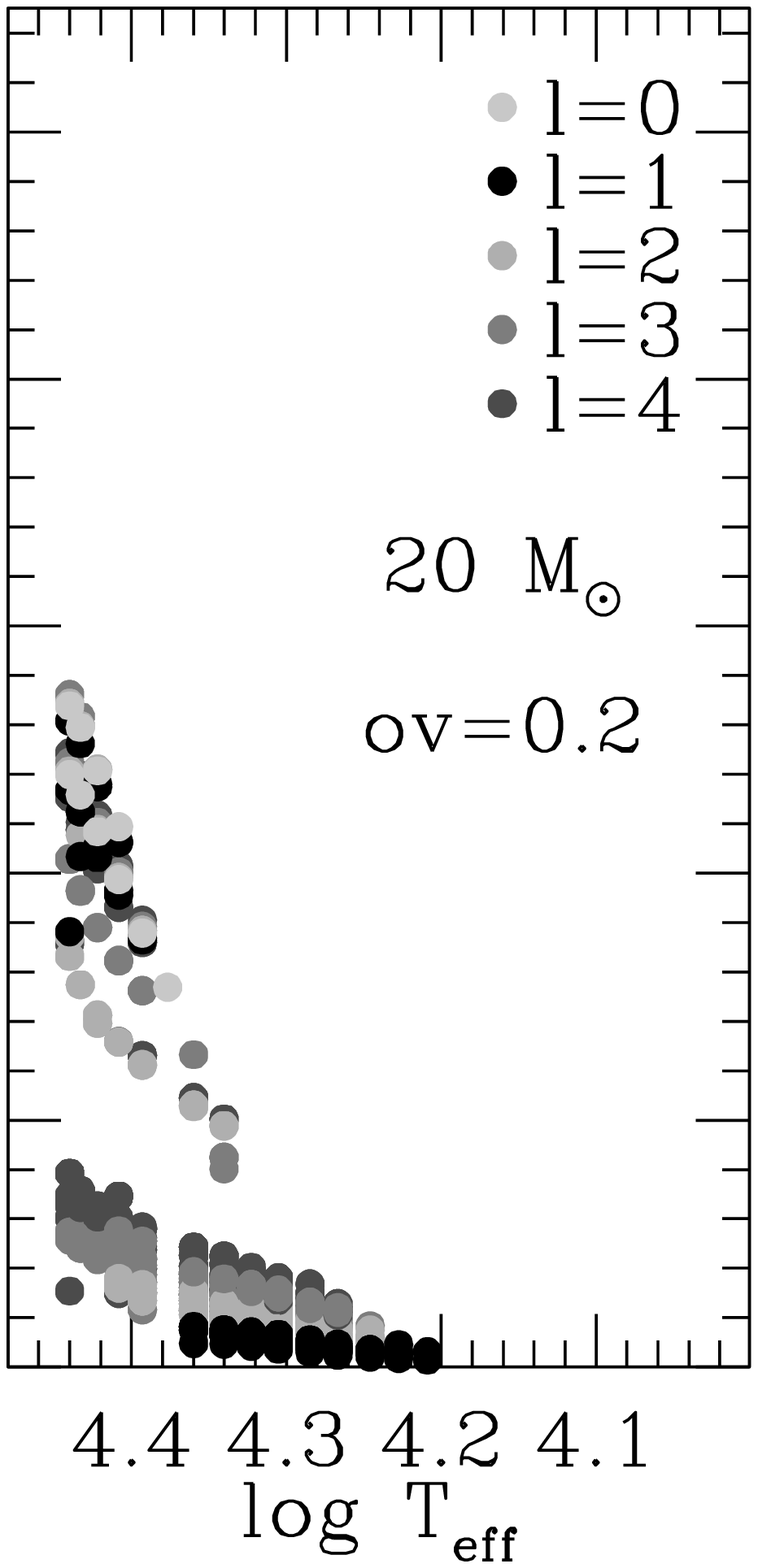}}

\caption{Excited g and p modes along the post-MS for $15\ M_{\odot}$ (left panel) and  $20\ M_{\odot}$ (right panel)  models based on Ledoux's criterion without (left) and with (right) overshooting.}
\label{osci1}
\end{figure}

The frequency range of excited modes and the $T_{\rm eff}$ domain of the instability strip are shown in Fig.\ref{osci1}.

\section*{Conclusions}

In models computed with the Schwarzschild's criterion, the ICZ which is closely related to the H-burning shell is located within the $\mu$-gradient region. In the ICZ, $N_{\rm BV}=0$ which corresponds to less radiative damping. On the other hand, in models computed with the Ledoux's criterion, the ICZ is thin and is located at higher values of $m/M$, at the base of the homogeneous region and therefore $N_{\rm BV}$ remains high in the $\mu$-gradient region which leads to more radiative damping. As a consequence more modes are excited in models computed with the Schwarzschild's criterion than in models computed with the Ledoux's one.

\acknowledgments{YL is grateful to the European Helio- and Asteroseismology Network HELAS for financial support. JM acknowledges financial support 
from the Prodex-ESA Contract Prodex 8 COROT (C90199).}

\References{
\rfr Chiosi, C. \& Maeder, A. 1986, ARA$\&$A, 24, 329
\rfr Chiosi, C., Bertelli, G., \& Bressan, A. 1992, ARA$\&$A, 30, 235
\rfr Morel, P. \& Lebreton, Y. 2008, ApSS, 316, 61
\rfr Saio, H., Kuschnig, R., Gautschy, A. et al. 2006, ApJ, 650, 1111
}

\end{document}